# Mineralogical Characteristics of Harmattan Dust Across Jos North Central and Potiskum North Earthern Cities of Nigeria.

Falaiye, O.A.[1] andAweda, F. O.[2*]
[1]Department of Physics, University of Ilorin, Ilorin, Nigeria.
*[2]Department of Physics and Solar Energy, Bowen University, Iwo, Osun State, Nigeria.
*Correspondence: francisaweda@gmail.com, aweda.francis@bowenuniversity.edu.ng

**ABSTRACTS:** The trace metals and mineralogical composition of harmattan dust carried out on the samples collected at Jos ((9º55'N, 8º55'E)) and Potiskum ((11º43'N, 11º02'E) as revealed by PIXE and AAS machine using clean Petri Dishes and Plastic bowls of 10 cm in diameter aimed on the characteristics of the mineralogical and elemental composition of the harmattan dust carried out in Nigeria. Thirteen trace elements, Na, K, Ca, Mg, Fe, Cd, Zn, Mn, Cu, Si, Al, Ti, and Zr were determined and their concentrations were evaluated in different proportion. Minerals such as Quartz [$SiO_2$], Corundum [$Al_2O_3$], Hematite [$Fe_2O_3$], Lime [CaO], Periclase [MgO], Rutile [$TiO_2$], Zincite [MnO], Bunsenite [NiO], Cuprite [$Cu_2O$], Zincite [ZnO], Baddeleyite [$ZrO_2$], Litharge [PbO], Monazite [$P_2O_5$], Montrodydite [HgO] and Petzite [$Au_2O_3$] were also determined in different concentrations. The particle weight of the sample for the residential and commercial areas were calculated to be Jos (18.95g/m$^2$, 19.25g/m$^2$), Potiskum (24.24 g/m$^2$, 2515g/m$^2$) respectively. The results shows that the harmattan dust that blows across the two stations in Nigeria comprise of high elements and more minerals.

## INTRODUCTION

An aerosol is a mixture of microscopic solid and liquid droplets suspended in air (Begum *et al.*, 2004). The chemical composition of fine aerosols poses adverse effects on human health especially in urban cities and plays vital role in climate change (Cheng et al., 2011; Sandhu et al., 2014; Tsai *et al.*, 2012). These particles can easily penetrate into the respiratory system (Chow et al., 1994; Pope III and Dockery, 2006) which may transfer some toxic element into the blood system. It has been reported that the chemical composition of fine aerosol has a negative impact on human health; it is however, uncertain (Pant and Harrison, 2013). As a result of dust that emanated from the Sahara, there are different types of ailment which causes different diseases during the period of harmattan dust. The harmattan period is usually associated with low and poor visibility of the atmosphere which is sometimes less than 1000m (Kalu, 1979; Falaiye *et al.*,

2003; Adimula *et al*., 2010; Falaiye *et al*., 2013). Falaiye *et al*., (2017) reported that the more the harmattan dust mass in the air the less the visibility of both human and animal. Studies have shown that harmattan season occurs between the months of November through March of the following year in Nigeria and during this period the atmosphere is laden with dust thus, reducing visibility and causing domestic as well as outdoor activities inconvenience (Aweda *et al*., 2017).

**MATERIALS AND METHODS**

As reported by Falaiye and Aweda (2018) clean plastic bowls of diameter 10 cm were exposed on an elevated platform and indoor of domestic buildings. Some of the bowls were exposed to collect dust particles for a period of one week, others for a period of one month. while some of the bowls were exposed to collect the dust samples for a period of five months (November to March) of the following year. A total of ten (10) samples were collected and stored in desiccators prior to the analysis in order to avoid contamination which could influence the results. This experiment follows what was done by (Falaiye, *et al*., 2013, Falaiye, *et al*., 2017) where Petri dishes were exposed on elevated platform to collect harmattan dust sample for mineralogical and chemical analysis to be conducted. The particle mass of the dust samples was also determined.

So also, clean glass slides surfaces rubbed with a thin film of vaseline and Whatman filter paper was employed to collect the dust sample (Direct Composition Method). Distilled water of about ten (10) liter volume was also exposed to air to collect the harmattan dust sample across the stations. During the collection process as reported by Falaiye, *et al*., (2013), measures such as keeping the samples containers away from public road and high ways were taken into consideration in order to minimize the input of the local dust.

The elemental and mineralogical composition analyses of the harmattan dust samples collected were carried out using the Absorption Atomic Spectroscopic (AAS) machine, Fourier Transform

Infrared (FTIR) Machine, and Particle Induced X-ray Emission (PIXE) were also used to determine the trace element analysis and mineralogical composition respectively of the harmattan dust samples collected.

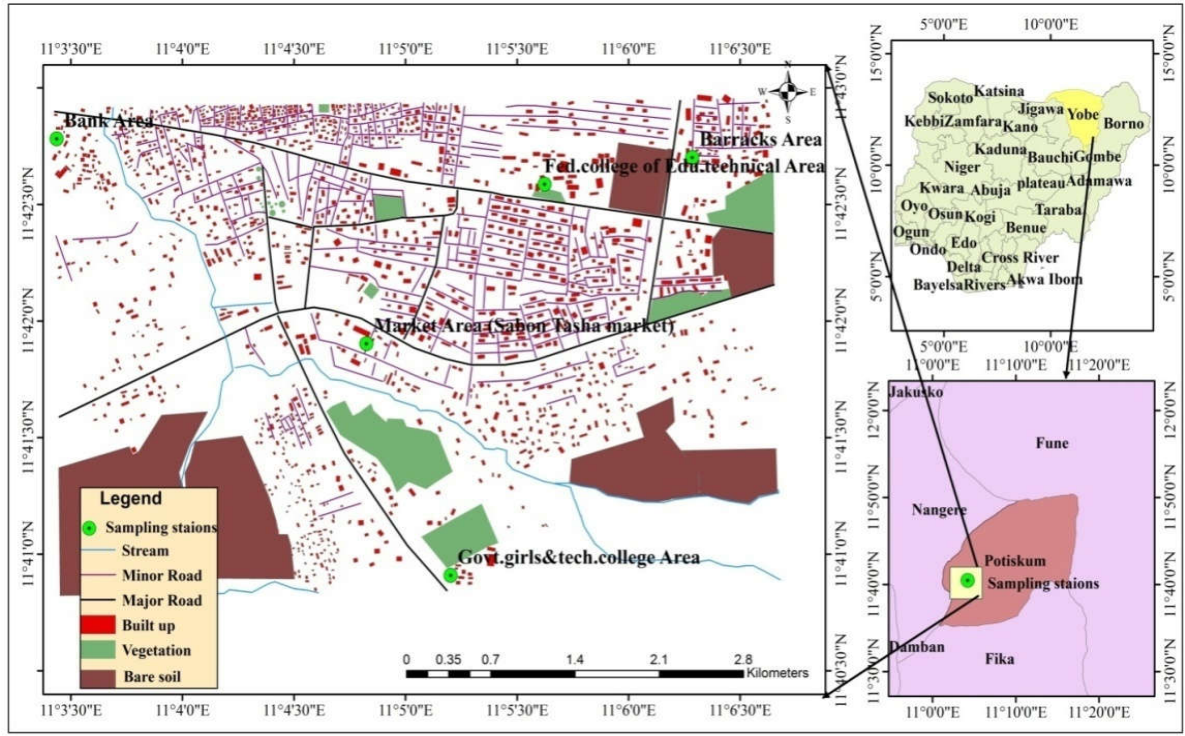

**Figure 1:** Map of Potiskum showing the sampling location

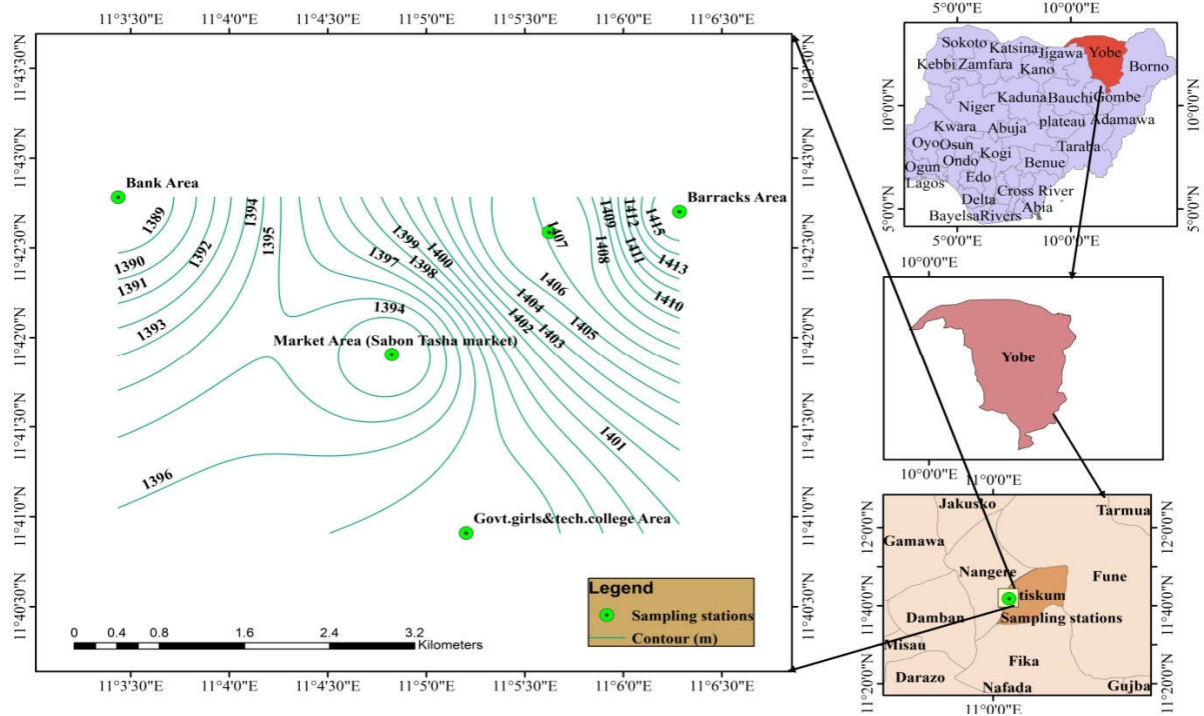

**Figure 2:** Map of Potiskum showing the contour line crossing the Sampling Location

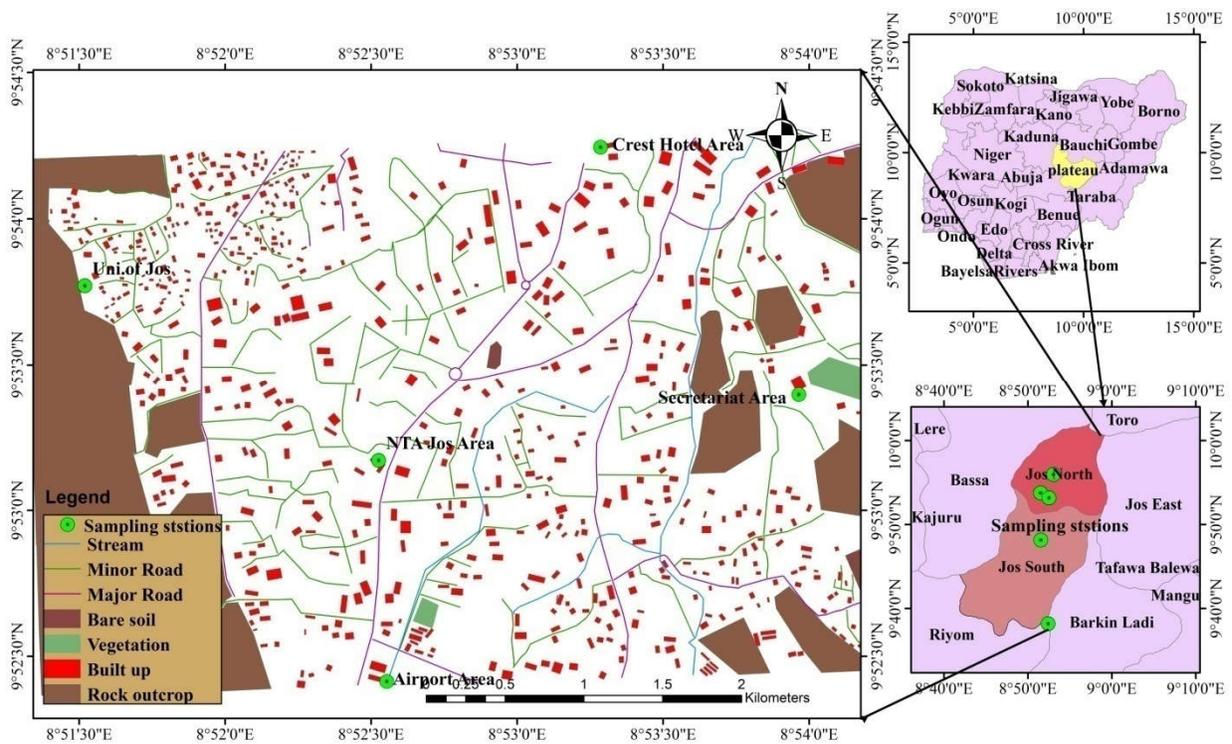

**Figure 3:** Map of Jos showing the sampling location

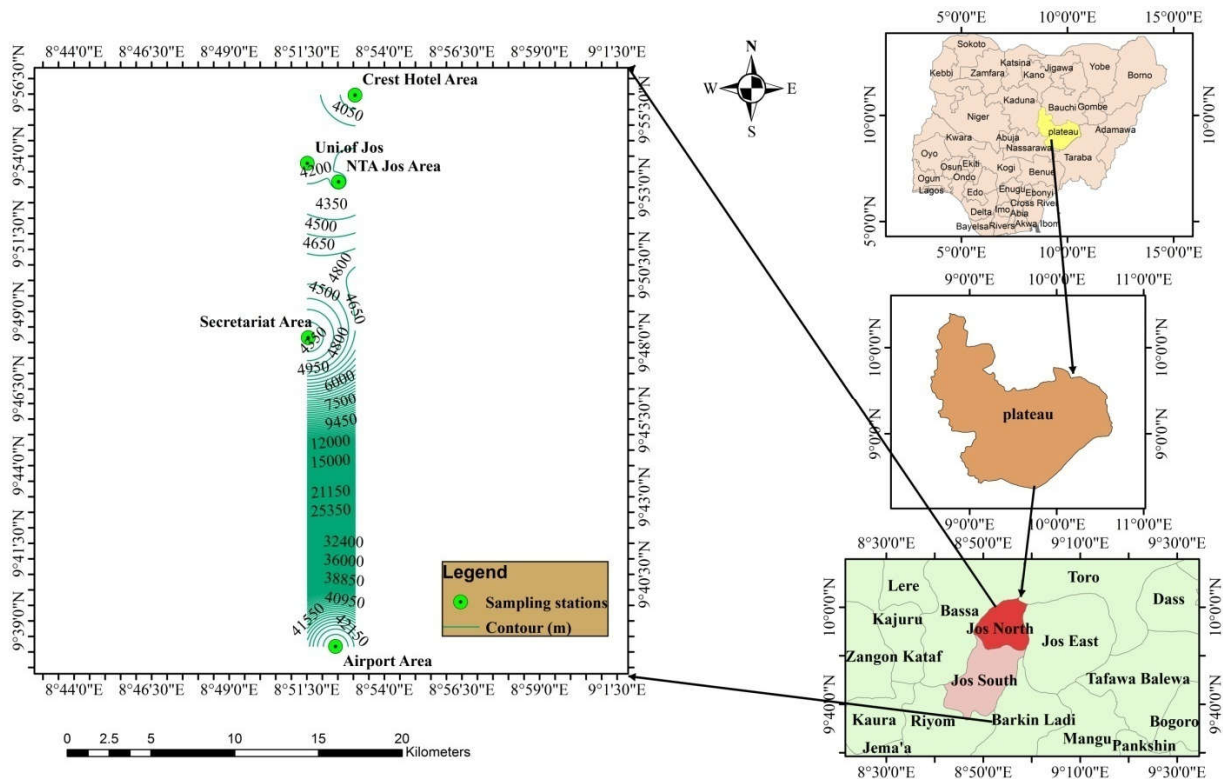

**Figure 4:** Map of Jos showing the contour line crossing the Sampling Location

**Figure 3.2:** Geographical Characteristics of the Stations

| Location | Latitude (⁰N) | Longitude (⁰E) | Altitude (m) | Average Rainfall (cm) | Climatic Classification | Vegetation |
|---|---|---|---|---|---|---|
| Potiskum | 11º43' | 11º02' | 508 | 713 | Tropical Continental | Sudan Savanna |
| Jos | 9º55' | 8º55' | 1217 | 1314.8 | Tropical Savanna | Guinea Savanna |

## RESULTS AND DISCUSSIONS

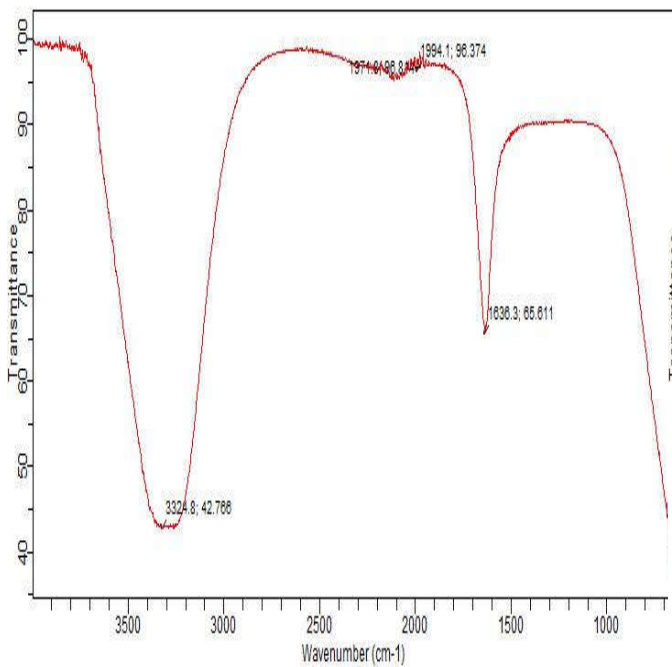 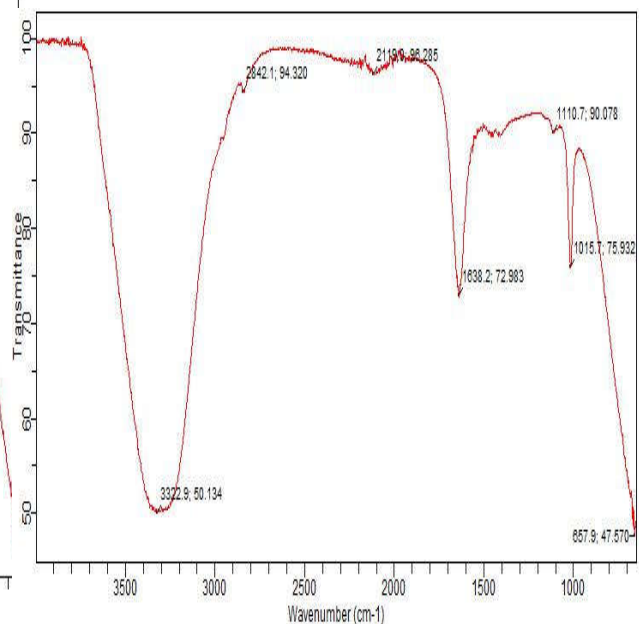

**Figure 4A:** A typical FTIR spectrum for Jos Liquid Harmattan Dust Sample

**Figure 4B:** A typical FTIR spectrum for Bauchi Liquid Harmattan Dust Sample

Figure 4 (A and B) shows the quantitative analyses carried out on the liquid samples collected at the stations considered for this research. This analysis was carried out to determine the major and minor constituent minerals present in the samples using the band position of the peaks from the prominent FTIR absorption peaks. These minerals were identified with the available instrument to determine the quantities present in the samples collected.

**Composition of Quartz Mineral**

From figure 4 (A and B) the FTIR absorption band appearing at 1638.2cm$^{-1}$ and 1015.7cm$^{-1}$ suggests the presence of quartz in the samples. The bending vibration at 1971.9cm$^{-1}$, symmetrical stretching vibration at 1992.3cm$^{-1}$ are assigned, the pattern of absorption in quartz can be explained by ascribing the 1971.9cm$^{-1}$ band region (Si-O asymmetrical bending vibration), the band region 1994.1cm$^{-1}$ (Si-O symmetrical bending vibrations), the bands in the region

1966.2cm$^{-1}$ (Si-O symmetrical stretching vibration). It was observed that there is maximum of four to six peaks in any of the sample collected across each location.

**Composition of Clay Minerals**

The presence of kaolinite, illite and montmorlinte it an indication of clay mineral in the samples collected across all these locations. Kaolinite is said to be clay mineral in crystallization which occurs in the monoclinic system and form the major constituent of Nigeria clay. It is also shown that harmattan dust in air reduces the visibility of air craft that may lead to air crash in some period because of the dusts that are lifted high as far as the stratospheric region of the atmosphere before dropping into the lower atmosphere after travelling to a very long distance.It can be observed from figure 4 (A and B) that the FTIR absorption peaks appearing at 1015.7cm$^{-1}$ in the sample indicate kaolinite. Absorbance at 1030cm$^{-1}$ is attributed to Si-O stretching of clay minerals.

Table 4.0: Mean concentrations of metals in different elemental form (ppm)

| Stations | Na (ppm) | K (ppm) | Ca (ppm) | Mg(ppm) | Fe(ppm) | Cd(ppm) | Zn(ppm) | Mn(ppm) | Cu(ppm) |
|---|---|---|---|---|---|---|---|---|---|
| Jos | 6.27 | 3.4 | 2.921 | 5.222 | 2.107 | 0.157 | -0.56 | -0.243 | -0.081 |
| Potiskum | 2.69 | 9.96 | 21.026 | 7.541 | 13.636 | 0.157 | 8.611 | 0.235 | -0.04 |

The elemental compositions and concentrations of the harmattan dust were determined in part per million (ppm) as revealed by AAS shown in the Table 4.0 above. The mean values of the elemental concentrations obtained for the period of harmattan season shows that element such as Na, K, Ca, Mg, Fe, Cd, Zn, Mn and Cu were presents in the samples collected across the two stations. These elements are in different concentrations. It was observed that Na for Jos and Potiskum were 5% and 2% respectively; the following observations were made K Jos(4%) and Potiskum(11%); Ca Jos(3%) and Potiskum(25%); Mg Jos(10%), and Potiskum(15%); Fe Jos(3%) and Potiskum(17%); Cd 10% across the two stations considered; for Zn Jos(-3%) and Potiskum(43%); For Mn Jos(-13%) and Potiskum(13%); For Cu Jos(-16%) and Potiskum(-8%). The aim of this research work is to establish the aerosol elemental composition present in the harmattan dust sample collected at each station. Noone *et al*., 2004 noted that aerosol directly influence climate by scattering or absorbing incoming solar radiation, and indirectly influence climate by acting as nuclei on which clouds can form. The present of Pb particulate matter in air are as a result of the activities taking place in the environment of the sample collection, this activity which includes fossil fuel combustion (including vehicles), metal processing industries and waste incineration around the sample collection area. Research as shown that the little amount of Pb can be harmful to human beings more especially little babies and young children as their body are tender to the absorption of high elements. If much of Pb is been inhale by a pregnant woman it may affect the health of the unborn child. Exposure has also been linked to impaired mental function, such as visual – motor performance and neurological damage in children, as well as memory and attention span The Air Quality Archive (2003). According to the Arizona Ambient Air Quality (1999), stated that the acceptable level for Fe, Ca, Mg and Zn in air for human respiration is 0.4 mg/m$^3$ for twenty four hours. Meanwhile, the concentration of this research result according to the values reveled by AAS machine with Jos (2107 mg/m$^3$) and

Potiskum (13636 mg/m$^3$). These show that the harmattan dust that blows across the two stations under consideration are far above the World Health Organization (WHO) recommended dosage of the Iron in human body. These could be as a result of the activities going on in all this environments such as vehicular movement and industrial activities that may be taking place in the area. Comparisons with the standard shows that the concentration of Iron in air for all the stations considered have the percentage values as follow Jos (3%) and Potiskum (17%). This shows that the percentage concentration of Potiskum is higher than Jos this may be as a result of Potiskum been close to the source of the dust that is the Sahara desert. This shows that the air across all the stations is not so clean due to little increase in the Iron content in all these stations. Meanwhile, Iron helps to keep plants and animals alive and plays major roles in the creation of chlorophyll in plants photo synthesis. More so, Iron is also an essential part of human and animal hemoglobin, the substance that carrier's oxygen within the red blood cells in the human being. But the excess of Iron in the body causes liver and heart diseases according to the WHO. From the above result it was observed that magnesium across the two stations considered having the value Jos (5222 mg/m$^3$) and Potiskum (7541 mg/m$^3$). This shows that the concentration of Potiskum is higher as compared with other location due to proximity to the source of the dust. This also shows that the magnesium content in air across all these stations are higher than the standard value recommended by World Health Organization (WHO). For calcium the acceptable level in air is observed to be the same as that of Iron. But for this research it was observed that the calcium content in air is far above the standard value as recommended by Word Health Organization (WHO). For manganese, this is use in the manufacture of cell batteries and as an oxidizing agent in some chemical industries. Has recommended, the Nigerian tolerable value of manganese was observed to be 0.01 mg/m$^3$. But for this research the Postiskum with the value 0.13 mg/m$^3$ while Jos with the value 0.11 mg/m$^3$. The present of Zinc (Zn) in air is essential in the elemental composition of harmattan dust because Zn is indispensable for human health and for all living organisms. Research has shown that Zn can be toxic to human health if the concentration is far higher than the WHO recommendations. It is can be observed that the acceptable level of Zn in air is the same has that of Iron. But for this research it was observed that the level of Zn in Potiskum and Minna shows a high value as compared with the WHO standard this may be due to some activities taking place at the two stations. But for station like Abuja, Lafia, Jos and Maiduri have negative value of Zn in air showing that they towns are very free from the effect of Zn. But for Iwo, it was observed that Iwo Zero concentration of Zn which also shows that the town is free from the concentration of Zn in air. The element potassium observed in the sample collected across each station under consideration shows that Iwo has the highest value of 25.97 mg/m$^3$, which was followed by Maiduguri and Minna with the value of 19.7 mg/m$^3$. This was then followed by Bauchi with the value 15.22 mg/m$^3$, while other stations have the value below 10 mg/m3. The acceptable value of potassium in air as recommended by WHO was observed to be 8.7 mg/m$^3$. Meanwhile, stations like Oyo, Ilorin, Abuja, Lafia, Jos and Postiskum are free from the toxic effect of potassium due to the fact that they have low concentration as been revealed by AAS machine. Therefore it shows that those stations with lower value of potassium are relatively clean in potassium. Potassium is a pent borate white, odorless, power substance which means that it is flammable, combustible, or explosive and dermal toxicity element. All these are in line with what was reported by Chineke and Chiemeka 2009.

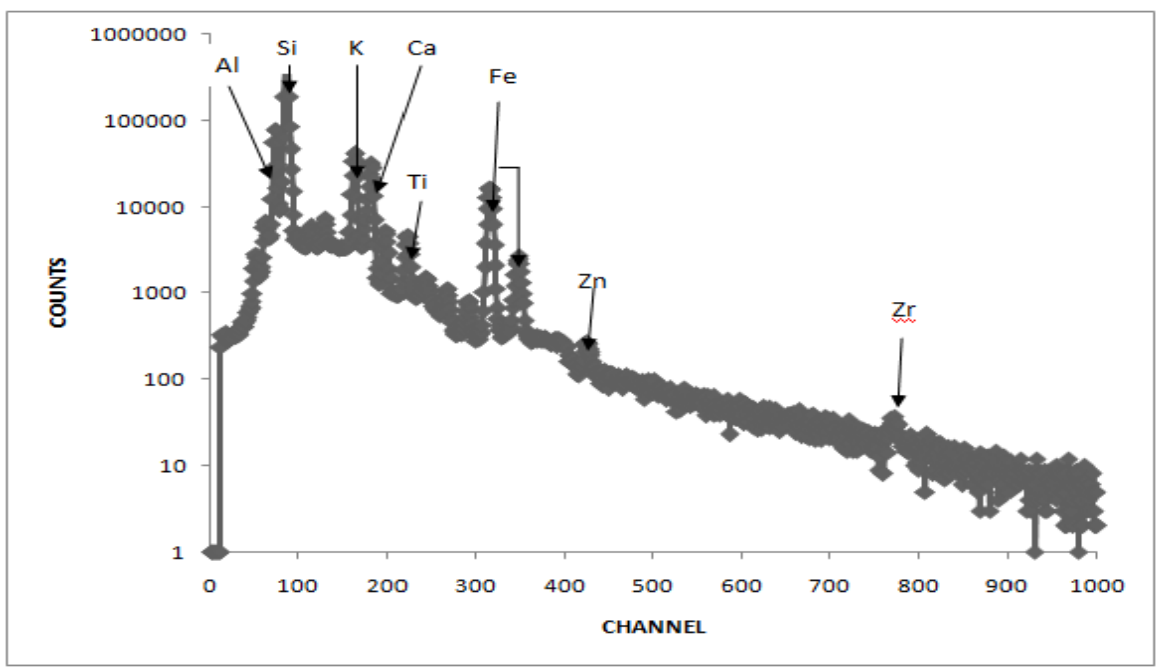

**Figure 5:** A PIXE Spectrum of Harmattan Dust Sample from Jos Deposit

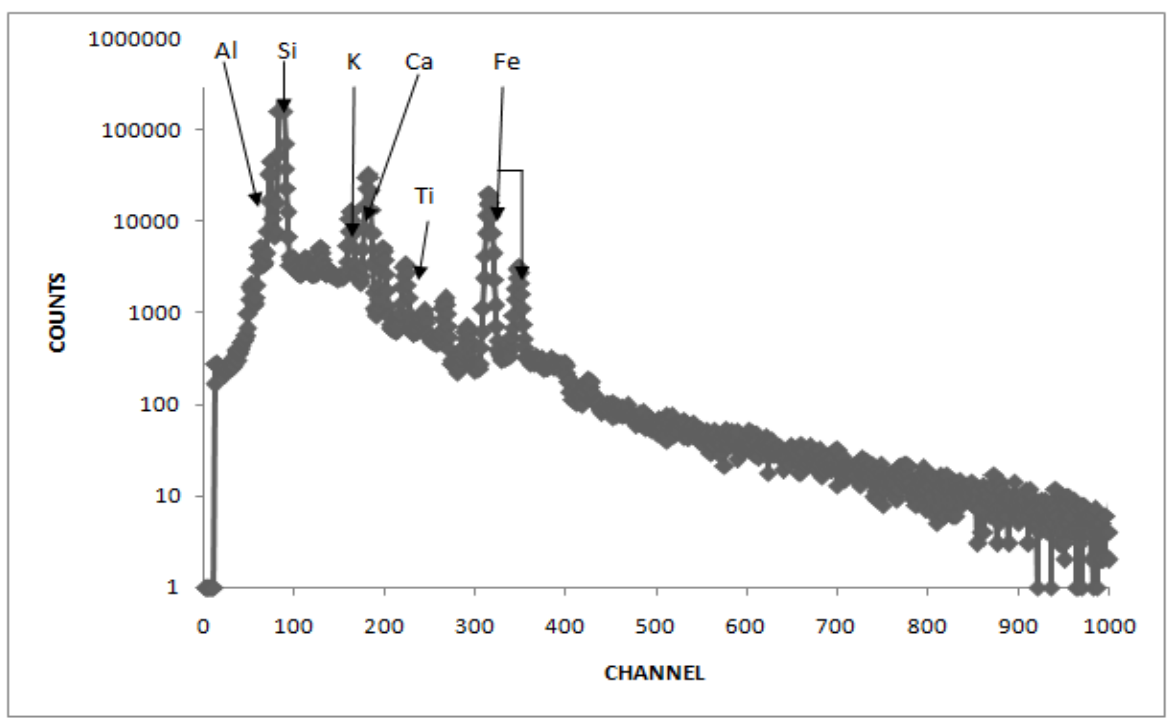

**Figure 6:** A PIXE Spectrum of Harmattan Dust Sample from Potiskum Deposit

**Mineralogical analysis of Harmattan Dust Using PIXE**

From the table above, it was observed that mineral such as Quartz [$SiO_2$](80%) with specific gravity 2.65, Corundum [$Al_2O_3$](6.94%) 4.0 - 4.2, Hematite [$Fe_2O_3$](3.84%) 5.26 and Lime [$CaO$](2.76%) 3.3. Its shows that Quartz is the dominant from the sample collected at Jos and Potiskum. While other minerals are present in small quantities or traces elements as shown in the table above. Minerals such as Periclase [$MgO$](0.68%) 3.56, Rutile [$TiO_2$](0.50%) 4.23, Zincite [$MnO$](0.07%) 5.66, Bunsenite [$NiO$](0.003%) 6.898, Cuprite [$Cu_2O$](0.013%) of 6.13 - 6.15, Zincite [$ZnO$](0.13%) 5.66, Baddeleyite [$ZrO_2$](0.54%) 5.4 – 6.02, Litharge [$PbO$](0.009%) 9.14, Monazite [$P_2O_5$] (0.2%) 4.6 – 5.4. Were minerals such as Montrodydite [$HgO$] 11.23 and Petzite [$Au_2O_3$] with specific gravity 9.13 were not detected that is, it has a zero percentage oxides. These minerals gave lower concentration from the sample collected at Jos and Potiskum.

It was also observed that oxides such as $Na_2O$ (0.69%), $SO_3$ (0.34%), Cl (0.34%), $K_2O$ (3.08%), $CrO_3$ (0.02%), $As_2O_5$ (0.004%), $Rb_2O$ (0.005%), SrO (0.004%) were also detected without any minerals that matches such oxide but other oxides such as $Sc_2O_3$, $V_2O_3$, $Ga_2O_3$, $Nb_2O_5$ and BaO have no percentage concentration which give such oxide without or no minerals present in them. If some of the oxides are combining they produce minerals such as Geikielite [$MgTiO_3$], Perovskite [$CaTiO_3$], Zinnmeta titanate [$ZnTiO_3$] and nickel titanium oxide [$NiTiO_3$]. These showed that the combination of $TiO_2$ produces some other minerals.

The study shows that minerals such as Quartz [$SiO_2$](80%) is the dominant constituent of the sample collected at the two stations (Jos and Potiskum). While other minerals present in small traces as shown in the table above. These results are in line with what was discovered in Ilorin and Ile-Ife by Falaiye et al., (2013) and Adedokun *et al*., (1989). With minerals such as

Corundum $[Al_2O_3]$(5.00%), Hematite $[Fe_2O_3]$(5.71%), and Lime $[CaO]$(3.39%). Minerals such as Periclase $[MgO]$(0.61%), Rutile $[TiO_2]$(0.45%), Zincite $[MnO]$(0.07%), Bunsenite $[NiO]$(0.006%), Cuprite $[Cu_2O]$(0.03%), Zincite $[ZnO]$(0.10%), Baddeleyite $[ZrO_2]$(0.03%), Monazite $[P_2O_5]$ (0.24%), Montrodydite $[HgO]$(0.01%) while Litharge $[PbO]$ were not present in the sample collected at Oyo. Petzite $[Au_2O_3]$(0.02%).

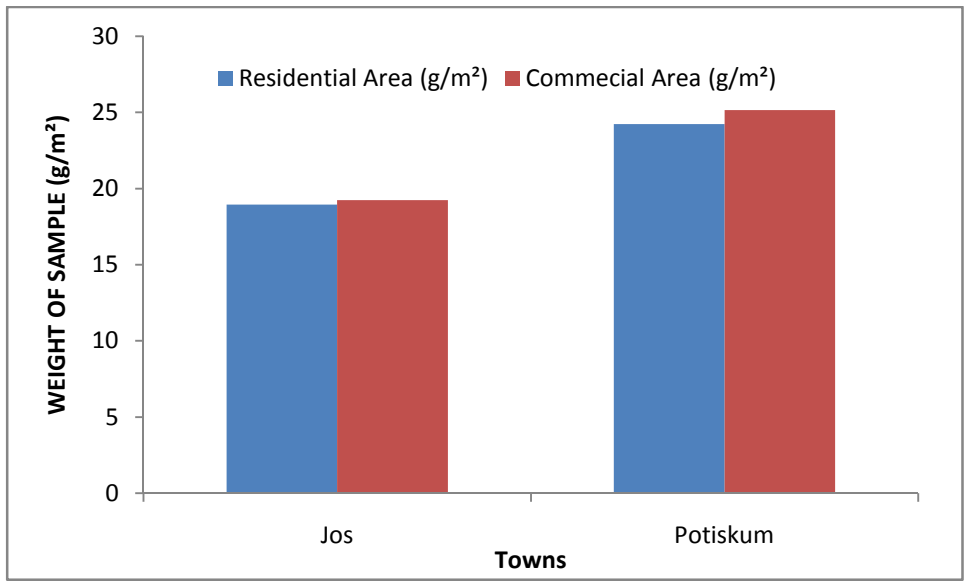

**Figure 7:** Average Mass of Dust Samples across each Station

**Harmattan Dust Mass across the Stations**

From figure 4.23 above, show the average mass of harmattan dust samples collected across each location. This shows that the residential area of Jos has the value 18.95 g/m² while the commercial areas have 19.25 g/m². For Potiskum, it was observed that the residential area have the value 25.15 g/m². According to Falaiye, *et al.*, (2017) depositions at higher elevations are higher because they would have become more dispersed by settling on vehicles, human beings and other objects which are more at lower elevation hence a lower concentration. These shows that harmattan dusts are particulate matter during it season. This could be as a result of intensity of human interaction such as vehicular movement, human movement trampling the ground, road side kitchens, open air waste incineration and wood burning (Anuforom *et al*., 2008), which may influence the weight of the dust at commercial areas. The result shows that the mass of the dust reduces along the trajectory part which could be as a result of wind transportation of the harmattan dust. The results of Jos present lower values as compared to what was observed in Potiskum which may be due to the dust settling along the way as it transversal from the source in the North eastern direction towards the Atlantic ocean as reported by Falaiye and Aweda (2018).

**CONCLUSION**

The abundance of quartz shows that the harmattan dust that blows across Nigeria has more of the mineral. The results showed that quartz percentage of Jos and Potiskum are higher than what was observed at Ilorin and Ile-Ife. This could be as a result of road construction taking place during the period of dust collection as a result of dust transportation in the air. It is also shown that harmattan dust in air reduces the visibility of air craft that may lead to air crash in some period because of the dusts that are lifted high as far as the stratospheric region of the atmosphere before dropping into the lower atmosphere after travelling to a very long distance. All these may have effect on the dust collected across the station since Jos and Potiskum are cities located in the country. In similar vein, the presence of elements in the sample gathered across Jos and Potiskum stations shows that the dusts, in Nigeria have almost all the elements present in harmattan. Some of the elements are in lower quantity and some are in high quantity. Elements such as Cu, Zn, Fe, Ca, Mn, Ni, As, K, Ti, Zr, and Cd and others are the major component of aerosol present in the harmattan sample collected across the two stations. Therefore, the toxic heavy metals like As, Ni, Zn, and Cu were found to be highly enriched around Jos and Potiskum. This could be as a result road construction during the period of the sample collection. The deposition rate and velocities of the major and minor elemental constituents of harmattan dust support the fact that harmattan dust is part of what reduces visibility and causes health effect during the period. However, the weight of the sample collected over Jos presents lower values as compared to locations further north of Nigeria. Meanwhile, Postiskum has higher percentage of the dust concentration as compared with Jos. This could be as a result of Potiskum closer to the source of the dust. More so, the dust weight is also being affected by vehicular movement around the towns under consideration and industrial smoke concentrated in some part of the city. Harmattan dust also causes some domestics disturbance as been reported by (Falaiye, 2008; Adefolalu, 1984; Adedokun *et al.,* 1989).The harmattan spells are often accompanied by droplets in the evening and early morning temperatures associated with an oscillation of the axis of the subtropical high (Adedokun, 1978; Adedayo, 1980; Adedokun *et al.,* 1989). This research was limited to five months (November to March) which are the major period of the harmattan dust in Nigeria.

Its shows that harmattan particulate dust matters have more weight if the sample collector is placed on the ground level which is half a meter than the sample collector placed at five meters high above the ground level. During the period of dust deposition (November- March), dust deposition around November to January have more deposition than from the period of February to March. This may be as a result of more aerosols which may be present during this period and also as a result of human interaction such as vehicular movement, human movement trampling the ground, road side kitchens, open air waste incineration and wood burning (Anuforom *et al*., 2008), in Ilorin during the period of sample collection. However, as compared to Nwadiogbu *et al.,* (2013) and Dimari *et al.,* (2008), it shows that the mass of harmattan dust in the Northern part have higher mass than that of the North Central towns of the country. This could be as a result wind transportation that blows the dust from the source and deposition along the trajectory path. The selection of the locations was random based on the vehicular traffic, population and free exposure of the environments. Therefore, it shows that the more the harmattan dust mass in the air, the less the visibility of both human and animal. However, as compared to Nwadiogbu *et al.,* (2013) and Dimari *et al.,* (2008), it shows that the mass of harmattan dust in the Northern part have higher mass than that of the North Central and South West towns of the country. This could be as a result wind transportation that blows the dust from the source and deposition along the trajectory path (Falaiye, *et al.,* 2017).